\documentstyle[12pt]{article}
\newcommand{\be}{\begin{equation}}
\newcommand{\ee}{\end{equation}}
\newcommand{\bea}{\begin{eqnarray}}
\newcommand{\eea}{\end{eqnarray}}

\newfont{\bg}{cmr10 scaled\magstep4}

\newcommand{\bigzerou}{\smash{\lower1.7ex\hbox{\bg 0}}}
\begin{document}
\title{
\begin{flushright} 
{\normalsize Gunma-Tech-98-02}
\end{flushright}
de Rham cohomology of ${\rm SO}(n)$ and some related manifolds by supersymmetric 
quantum mechanics }
\author{Kazuto Oshima \\ \\
\sl Gunma College of Technology,Maebashi 371- 0845,Japan \\}
\date{ }
\maketitle
\abstract{We study supersymmetric quantum mechanics on ${\rm RP}_{n},{\rm SO(n)},
{\rm G}_{2}$ and ${\rm U}(2)$ to examine Witten's Morse theory concretely. We confirm the simple instanton picture of the de Rham cohomology that has been given in a previous paper.  We use a  reflection symmetry of each theory to select the true vacuums. The number of  selected  vacuums agrees with the de Rham cohomology  
for each of the above manifolds.}
\maketitle
\\ 
\baselineskip=24pt
\newpage
\section{INTRODUCTION}
In a previous paper,${^1}$ the author has investigated Witten's
Morse theory${^2}$ for ${\rm SO}(n)$.$^{3}$ Instanton effects between
adjacent classical vacuums have been studied. Using reflection symmetries 
of the theory,  a selection rule for true vacuums has been given. 
This selection rule works
well, at least for $n \le 5$;$^{1}$  the number of selected vacuums 
agrees with the de Rham cohomology of ${\rm SO}(n)$. The main purpose of this paper
is to show that the selection rule works well for arbitrary $n$.
We also apply  the selection rule 
to some related manifolds ${\rm RP}_{n},{\rm G}_{2}$ and ${\rm U}(2)$
;${\rm RP}_{1} \simeq {\rm SO}(2)$,
${\rm RP}_{3} \simeq {\rm SO}(3)$,${\rm G}_{2} \subset {\rm O}(7)$ and
${\rm U}(2)  \simeq {\rm SO}(2) \times {\rm SU}(2)$.

This paper is organized as follows:
In Sec.II we review  supersymmetric quantum mechanics on a manifold  $M$. The selection rule for true vacuums is stated in this section. 
The manifold ${\rm RP}_{n}$ are studied in the third section. 
The manifold ${\rm RP}_{n}$ are simple and this section will be
helpful to understand
the following sections. In Sec.IV the de Rham cohomology of ${\rm SO}(n)$ is 
derived for arbitrary $n$ by the selection rule. The manifolds
${\rm G}_{2}$ and ${\rm U}(2)$ are discussed in Sec.V  and VI, respectively.
The results are summarized in the last section.
\section{SUPERSYMMETRIC  QUANTUM  MECHANICS ON A MANIFOLD}
The  supersymmetric hamiltonian on a manifold with Morse function
$h$ is given by 
\begin{equation}
\hat{H}=-{1 \over 2}(d_{h} d_{h}^{\dagger} +d_{h}^{\dagger} d_{h}),
\end{equation}
where $d_{h}=e^{-h}de^{h}, d_{h}^{\dagger}=e^{h}d^{\dagger}e^{-h},
d$ is the exterior derivative and $d^{\dagger}$ is its adjoint operator.
In (any) coordinetes $\{x^{\mu}\}$, the exterior multiplication $e_{dx ^{\mu}}$ and the interior multiplication
$i_{\partial \over \partial x^{\mu}}$ can be identified with the fermion creation
operator $\hat{\psi}^{* \mu}$ and the annihilation operator 
$\hat{\psi} _{\mu}$ and we have
\begin{equation}
d=\hat{\psi}^{* \mu} \nabla _{\mu}, \quad  
d^{\dagger}=g^{-{1 \over 2}}\nabla _{\mu} g^{1 \over 2}g^{\mu \nu}\hat{\psi} _{\nu},
\end{equation}
where $g$ is the determinant of the metric tensor $g_{\mu \nu}$
and $\nabla _{\mu}$ is the covariant derivative
\begin{equation}
\nabla _{\mu}={\partial \over \partial x^{\mu}}
-\Gamma_{\mu \nu}^{\lambda}\hat{\psi} ^{* \nu}\hat{\psi}_{\lambda}. 
\end{equation}
The hamiltonian $\hat{H}$ $[(1)]$ takes the form:
\begin{equation}
2\hat{H} =-g^{-{1 \over 2}} \nabla _{\mu}g^{1 \over 2}g^{\mu \nu} \nabla _{\nu}
+R_{\mu \nu \sigma \tau}
\hat{\psi}^{\sigma}\hat{\psi}^{* \tau}\hat{\psi}^{* \nu}\hat{\psi} ^{\mu}
+g^{\mu \nu} {\partial h \over \partial x^{\mu}}{\partial h \over \partial x^{\nu}}
+H_{\mu \nu}[\hat{\psi} ^{* \mu},\hat{\psi}^{\nu}],
\end{equation}
where  $R_{\mu \nu \sigma \tau}$ is the Riemann tensor, and
 $H_{\mu \nu}$ is the Hessian matrix
\begin{equation}
H_{\mu \nu}=(\partial_{\mu}\partial_{\nu}-\Gamma_{\mu \nu}^{\lambda}\partial_{\lambda})h.
\end{equation}
The corresponding Lagrangian is 
\bea
{\cal L} ={1 \over 2}g_{\mu \nu}{dx^{\mu} \over dt}{dx^{\nu} \over dt}
+{1 \over 2}g^{\mu \nu}{\partial h \over \partial x^{\mu}}
{\partial h \over \partial x^{\nu}}
+\psi ^{* \mu}({d \over dt}\psi_{\mu}-\Gamma_{\mu \nu}^{\lambda}\psi_{\lambda}
{dx^{\nu} \over dt}) \nonumber  \\ +H_{\mu \nu}\psi ^{* \nu}\psi^{\mu} 
+{1\over4}R_{\mu \nu \sigma \tau}\psi^{\mu}\psi^{\nu}\psi^{* \sigma}\psi^{* \tau}.
\eea
The gradient flow equation of (6) is
\begin{equation}
{dx^{\mu} \over dt}=\pm g^{\mu \nu}{\partial h \over \partial x^{\nu}}.
\end{equation}
A relevant instanton solution satisfies (7) and connects adjacent critical points.

There is one classical vacuum for each critical point of $h$.$^{2}$
Around each critical point, $h$ has the expansion 
\be
h= h_{0} +\sum _{i}\lambda _{i}\xi_{i}^{2},
\ee
where $\lambda _{i}$ are eigenvalues of the Hessian matrix and
$\xi_{i}$ are the corresponding local coordinates. The classical
vacuum around a critical point corresponds to an $l-$form:
\be
|l\rangle=\prod _{\lambda _{i} < 0} {\hat \psi}_{\xi _{i}}^{*}
|0 \rangle,
\ee
where $l$ represents the number of the excited fermions(the negative 
eigenvalues) and $|0 \rangle $ is the bosonic vacuum.  

The true vacuums are determined by the quantum tunneling
between adjacent classical vacuums (9).
According to Witten,$^{2}$ one has
\be
\langle l+1|d_{h}|l \rangle= \sum_{\gamma} n_{\gamma} e^{-(h(P^{(l+1)})-h(P^{(l)}))},
\end{equation}
where $n_{\gamma}$ is an integer assigned for each instanton path 
$\gamma$ from $P^{(l)}$ to $P^{(l+1)}$. If a state $|l \rangle$ does not couple with any adjacent classical
vacuums, that is if  $d _{h}|l \rangle =0$ and $\langle l |d _{h}=0$ , $|l \rangle $ is a true vacuum. 

If there are plural instanton paths connecting a pair of 
adjacent critical points,
it is not so easy to determine whether the matrix element (10) is zero or not,
because of the notorious minus signs associated with fermions. In fact, for 
${\rm SO}(n)$, there are exactly two instanton solutions between each pair of adjacent
classical vacuums.$^{1}$ The selection rule given in Ref.1 is as
follows. Find a reflection transformation that leaves $h,d_{h},
d_{h}^{\dagger}$, and consecuently $\hat{H}$ invariant and that interchanges the
two instanton paths. If the parities of a pair of adjacent classical vacuums are
the same, the corresponding matrix element does not vanish, and the  
two classical vacuums are not true vacuums. If the parities of the two 
classical vacuums are different, the corresponding matrix element 
vanishes, and the  two classical vacuums can be true vacuums.
Repeat this for each pair of adjacent classical vacuums, and
one obtains all true vacuums.

In the following sections, we apply this selection rule to ${\rm SO}(n)$ and
some related manifolds . We see that the number of true vacuums
agrees with the de Rham cohomology for each manifold.

\section{${\rm RP}_{n}$}
In this section, we discuss the real projective space ${\rm RP}_{n}$,
which is very simple but non-trivial. The manifold  ${\rm RP}_{n}$ can be
identified with $(n+1) \times (n+1)$ real matrices $A$ with conditions
$A^{t}=A$,$A^{2}=A$ and ${\rm tr} A=1$.$^{4}$  Concretely, we have
\be
A=( x_{i}x_{j}) , \quad i,j=0,1,2,\cdots,n,
\ee
where $\displaystyle \sum _{i=1}^{n}x_{i}^{2} \le 1$ and 
$x_{0}=\sqrt{1- {\displaystyle \sum _{i=1}^{n}x_{i}^{2}} }$. 
Let us introduce the following polar coordinates
\bea
x_{1}&=&\sin \theta _{1} \cos \theta _{2}, \nonumber \\
x_{2}&=&\sin \theta _{1} \sin \theta _{2} \cos \theta _{3},\nonumber \\
    \vdots   \nonumber \\
x_{n-2}&=&\sin \theta _{1} \sin \theta _{2} \cdots  
\sin \theta _{n-2}\cos \theta _{n-1}, \nonumber \\
x_{n-1}&=&\sin \theta _{1} \sin \theta _{2} \cdots  
\sin \theta _{n-2}\sin \theta _{n-1}\cos \theta _{n}, \nonumber \\
x_{n}&=&\sin \theta _{1} \sin \theta _{2} \cdots  
\sin \theta _{n-2}\sin \theta _{n-1}\sin \theta _{n},
\eea
with $-{\pi \over 2} \le \theta _{i} \le {\pi \over 2}$.

Fix real numbers $c_{0},c_{1},\cdots , c_{n}$ with 
$c_{i+1} > 2c_{i} > 0$. Then 
\be
h= \sum _{i=0}^{n}c_{i}x_{i}^{2}
\ee
is a Morse function on ${\rm RP}_{n}$$^{4}$ with critical points $P^{(l)}$     
\be
 P^{(l)} = {\rm diag}( 0, 0, \cdots, 0,x_{l}^{2}=1, 0, \cdots, 0 ).
\ee
Around $P^{(l)}$, $h$ has the expansion
\bea
h&=&(c_{0}-c_{l})(x_{0}x_{l})^{2}
+ \cdots+(c_{l-1}-c_{l})(x_{l-1}x_{l})^{2}+c_{l} \nonumber \\ 
&+&(c_{l+1}-c_{l})(x_{l+1}x_{l})^{2}+\cdots
+(c_{n}-c_{l})(x_{n}x_{l})^{2}.
\eea
The classical vacuum around $P^{(l)}$ is $l$ fermions excited state
\be
|l\rangle= {\hat \psi}_{0l}^{*}{\hat \psi}_{1l}^{*} \cdots 
{\hat \psi}_{l-1l}^{*}|0 \rangle,
\ee
where ${\hat \psi}_{ij}^{*}$ is the fermionic mode corresponding to 
$x_{i}x_{j}$.

We introduce  a metric on ${\rm RP}_{n}$ and find reflection symmetries of the
theory and instanton solutions concretely. We introduce the following metric
\be
g_{ij}={1 \over 2} {\rm tr}{\partial A^{t}  \over \partial {\theta}_{i}}
{\partial A \over \partial {\theta}_{j}}.
\ee
The non-zero components of $g_{ij}$ and Christoffel symbols are
\bea
g_{11}=1,g_{ii}=\sin ^{2}\theta _{1}\sin ^{2}\theta _{2} \cdots
\sin ^{2}\theta _{i-1} \qquad ,i=2, \cdots ,n \nonumber  \\
\Gamma _{ij}^{j}=\cot \theta _{i} \qquad ,i=1, \cdots ,n-1 , \quad j=i+1, \cdots , n.
\eea

The Morse function $h$ [(13)], the metric $g_{ij}$ and  
the corresponding supercharge
\be
d=\hat{\psi}^{* i} \nabla _{i}= \hat{\psi}^{* i}
({\partial \over \partial \theta _{i} }
-\cot \theta _{i} \sum _{j=i+1}^{n}\hat{\psi}^{* j} \hat{\psi}_{j}),
\ee
are invariant under the following transformations 
\be
[i] :\qquad \theta _{i},{\hat\psi}^{*i},{\hat \psi}_{i} 
\rightarrow -\theta _{i},-{\hat\psi}^{*i},-{\hat \psi}_{i}  \quad 
, i=1, \cdots ,n.
\ee
Subsequently, the theory is invariant under these reflection transformations. 
Under the transformation $[i]$, $x_{j}$ transform to  $-x_{j}$ if $j \geq i$, and
$x_{j}$ are invariant if $j < i$.  The fermionic  modes ${\hat \psi}_{jk}^{*}$
have the same transformation properties as $x_{j}x_{k}$

From the gradient flow equation (6), we have 
\bea
{d\theta _{i} \over dt} &=&(c_{0}-c_{i-1}+(c_{i}-c_{0})\cos ^{2}\theta _{i+1} 
+(c_{i+1}-c_{0})\sin ^{2}\theta _{i+1}\cos ^{2}\theta _{i+2} \nonumber \\
&+& \cdots
 +(c_{n}-c_{0})\sin ^{2}\theta _{i+1}\sin ^{2}\theta _{i+2} \cdots 
\sin ^{2}\theta _{n} )\sin 2\theta _{i}, \quad i=1, \cdots n-1 , \nonumber  \\
{d\theta _{n} \over dt} &=&(c_{n}-c_{n-1})\sin 2\theta _{n}.
\eea
One finds that there are exactly a pair of instanton paths that connect $|l-1\rangle $ and  
$|l\rangle $ [(16)]
\bea
A=
\bordermatrix{
    &    &    &    & l-1 & l &   &   &  \cr
    & 0  &    &    &     &   &   &   &   \cr
    &   & \ddots   &    &     &   &   &   &  \cr
    &   &    & 0  &     &   &   &   &  \cr
l-1 &   &    &    & \cos ^{2}\theta _{l} & \pm \sin \theta _{l} \cos \theta _{l}
                              &   &    &   \cr
l   &   &    &    & \pm \sin \theta _{l} \cos \theta _{l} & \cos ^{2}\theta _{l}
                              &   &    &   \cr
    &   &    &     &    &      & 0 &    &   \cr
   &   &    &     &     &     &   &  \ddots   &   \cr
  &   &    &     &      &    &   &       & 0  \cr
},
\eea
where $0 \leq \theta _{l} \leq {\pi \over 2}$.
Each of these two instanton solutions causes non-zero instanton effect between
$|l-1\rangle $ and  $|l\rangle $.

Let us consider the transformation $[l]$. This transformation interchanges
the two instanton paths. Under this transformation the two classical vacuums
transform as
\be
|l-1\rangle \rightarrow |l-1\rangle ,\qquad 
|l\rangle \rightarrow (-1)^{l}|l\rangle .
\ee
Thus if $l$ is even, the two classical vacuums $|l-1\rangle $ and  
$|l\rangle $ have the same parities, which means that
the classical vacuums $|l-1\rangle $ and $|l\rangle $ are quantum mechanically
coupled, and are not true vacuums.$^{1}$  If $l$ is odd, the 
two classical vacuums $|l-1\rangle $ and  $|l\rangle $ have the 
opposite parities and the two instanton effects cancel each other. 

One obtains the following results.  For $n=$even there is only one true vacuum $|0 \rangle $,
and for $n=$odd there are two true vacuums  $|0 \rangle $ and $|n \rangle $.
These results agrees with the de  Rham cohomology of ${\rm RP}_{n}$;for $n=$even,
there is one non-zero betti number $b_{0}=1$, and for $n=$odd, there are two
non-zero betti numbers $ b_{0}= b_{n}=1$.

\section{${\rm SO}(n)$}
In Ref.1, all the classical vacuums, all the instanton solutions and symmetry transformations are identified for a certain Morse function on ${\rm SO}(n)$. 
We review them to find true vacuums for arbitrary $n$.

Let $A=(a_{ij})$ be a group element of ${\rm SO}(n)$, and fix real numbers $c_{1},c_{2},...,c_{n}$ with $c_{i}>2c_{i+1}>0$.
Then
\be
h=\sum _{i=1 }^{n} c_{i}a_{ii} ,
\ee
is a Morse function$^{4}$ on ${\rm SO}(n)$ with critical points $P^{(l)}$, 
\be
P^{(l)}={\rm diag}(\epsilon _{1}, \epsilon _{2}, \cdots ,
\epsilon _{n}), \qquad \epsilon _{i}=\pm 1, \prod _{i} \epsilon _{i}=1.
\ee
The Morse index of $P^{(l)}$ is ${\displaystyle \sum _{i=1}^{n}}
(n-i){1+{\epsilon _{i}} \over 2}$. 
Around each critical point, $h$ has the expansion 
\be
h=\sum _{i} \epsilon _{i}c_{i} +\sum _{i<j}(\lambda _{ij}\xi_{ij}^{2}+
\mu _{ij}\eta _{ij}^{2}),
\ee
where
\be
\lambda _{ij}=-{\frac{\epsilon _{j}-\epsilon _{i}} 4}(c_{j}-c_{i}),
\qquad \xi_{ij}  = a_{ij}+a_{ji},             
\ee 
\be
\mu _{ij} =-{\frac{\epsilon _{j}+\epsilon _{i}} 4}(c_{j}+c_{i}),
\qquad \eta _{ij}=a_{ij}-a_{ji}.          
\ee

A vacuum state which is localized around a critical point with the Morse index
 $l$ is
\be
|l\rangle=\prod _{i < j} \prod _{\epsilon _{i}=\epsilon _{j}=1}{\hat \psi}_{\eta _{ij}}^{*}
\prod _{\epsilon _{i} >\epsilon _{j}}{\hat \psi}_{\xi _{ij}}^{*}|0 \rangle .
\ee

There are exactly a pair of instanton solutions between the critical  points
${\rm diag}(\epsilon _{1}, \epsilon _{2}, \\ \cdots ,\epsilon _{i-1},-1,1,
\epsilon _{i+2},\cdots, 
\epsilon _{n})$ and ${\rm diag}(\epsilon _{1}, \epsilon _{2}, \cdots ,\epsilon _{i-1},1,-1,
\epsilon _{i+2},\cdots, 
\epsilon _{n})$ :
\be
\left(
\begin{array}{cccccccc}
\epsilon _{1} & & & & & & &  \nonumber \\
            & \ddots & & & & & & \nonumber \\
            &   & \epsilon _{i-1} & & & & & \nonumber  \\
            &   &  & \cos \theta & \mp \sin \theta & & & \nonumber \\
    &   &  & \mp  \sin \theta & - \cos \theta & & & \nonumber \\
            &   &  &   &    & \epsilon _{i+2}   &    & \nonumber \\
            &   &  &   &    &    & \ddots   & \nonumber \\
            &   &  &   &    &    &    &  \epsilon _{n}  \\
\end{array}
\right).\\
\ee
Some changes of the sings are needed in (30) for the instanton solutions 
between the critical points ${\rm diag}(\epsilon _{1}, \epsilon _{2}, \cdots , 
\epsilon _{n-2},-1,-1)$ and ${\rm diag}(\epsilon _{1}, \epsilon _{2}, \cdots , 
\epsilon _{n-2},1,1)$. 

The theory is invariant under each of the supersymmetric 
generalizations$^{1}$ of  the following  reflection transformations  
\be
a _{ij} \rightarrow -a _{ij} \quad {\rm and} \quad  a_{ji}\rightarrow -a_{ji} \quad , j \ne i ,i=2,3, \cdots, n.
\ee
and  the transposition    
\be
a _{ij} \rightarrow a _{ji} .
\ee
We call the corresponding supersymmetric transformations as $[i]$ and
$[t]$, respectively.
Supersymmetric generalization means that the corresponding fermionic
modes are also transformed. A transformation that is useful to elucidate
true vacuums is the transformation that interchanges the two instanton
solutions (30). 

For the instanton process corresponding to (30), the transformation $[i]$ is relevant.  From (29), one sees what kind of fermionic
modes are newly exited or suppressed in this process.  In this process,
only  the fermionic modes including the indices $i$ or $i+1$ can newly be exited or suppressed. 
At first, the fermionic modes
 ${\hat \psi}_{ki}^{*}$ and ${\hat \psi}_{ki+1}^{*}, k=1, \cdots ,i-1$,
are excited  in  pair
$ {\hat \psi}_{\xi _{ki}}^{*}{\hat \psi}_{\eta _{ki+1}}^{*}$
if $\epsilon _{k}=1$. 
They change into   $ {\hat \psi}_{\eta _{ki}}^{*}{\hat \psi}_{\xi _{ki+1}}^{*}$ in this process, but do not cause any changes in their parities.
In this process, the mode ${\hat \psi}_{\xi _{ii+1}}$ is newly exited. 
As for the fermionic modes
 ${\hat \psi}_{ik}^{*}$ and ${\hat \psi}_{i+1k}^{*}, k=i+1, \cdots ,n$,
 the corresponding exited modes are
  ${\hat \psi}_{\eta _{i+1k}}^{*}$ and ${\hat \psi}_{\xi _{i+1k}}^{*}$, 
for  $\epsilon _{k} =1$ and $\epsilon _{k}=-1$, respectively. 
In this process they  change into ${\hat \psi}_{\eta _{ik}}^{*}$
and  ${\hat \psi}_{\xi _{ik}}^{*}$,respectively. Thus, in this process
the number of the index $i$ increases by $n-i$, and the number of the 
index $i+1$ diminishes by $n-i-2$ in the exited fermionic modes. The number
of exited $\eta$-modes does not change, so this process is not prohibited
by the transformation $[t]$( the instanton process below (30) is prohibited
by  $[t]$). Taking the transformation $[i]$
or $[i+1]$ into consideration, we see that the two instanton effects cancel
each other if $n-i$ is odd, and they add up if $n-i$ is even.  
Applying this simple rule to all the instanton solutions, one can easily 
select all true vacuums.

Let us first consider ${\rm SO}(2n-1)$. We mark the elements of the critical
points off two by two
\be
{\rm diag}(\epsilon _{1}, \epsilon _{2}
\mid \epsilon _{3}, \epsilon _{4} \mid \cdots 
\mid \epsilon _{2n-3},\epsilon _{2n-2} \mid \epsilon _{2n-1}).
\ee 
From the above rule, one sees that there are instanton effects between the two states around the following critical points  
\be
{\rm diag}(\epsilon _{1}, \epsilon _{2}
\mid \cdots  \mid 1, - 1 \mid \cdots \mid \epsilon _{2n-1})
\leftrightarrow
{\rm diag}(\epsilon _{1}, \epsilon _{2}
\mid \cdots  \mid -1,  1 \mid \cdots \mid \epsilon _{2n-1}).
\ee 
On the contrary, if $\epsilon_{2i-1}=\epsilon_{2i}(i=1,2, \cdots n-1)$ ,
the corresponding state does not affected by instanton effects, and 
it is decided to be a true vacuum. Thus, the critical points that correspond
to true vacuums are
\be
{\rm diag}(\epsilon _{1}, \epsilon _{1}
\mid \epsilon _{3}, \epsilon _{3} \mid  \cdots 
\mid \epsilon _{2n-3}, \epsilon _{2n-3} \mid 1).
\ee Their Morse indices are  $\displaystyle \sum _{i={\rm odd}}^{2n-3}(4n-3-2i)\epsilon _{i}$.
This formula 
agrees with the de Rham cohomology of ${\rm SO}(2n-1)$:$\Lambda (x_{3},x_{7},\cdots,
x_{4n-5})$.$^{5}$

In the same way, for  ${\rm SO}(2n)$, the critical points that correspond to true vacuums are found to be 
\be
{\rm diag}(\epsilon _{2n-1} \mid \epsilon _{1}, \epsilon _{1}
\mid \epsilon _{3}, \epsilon _{3} \mid  \cdots 
\mid \epsilon _{2n-3}, \epsilon _{2n-3} \mid \epsilon _{2n-1} ).
\ee 
Their Morse indices are  $\displaystyle \sum_{i={\rm odd}}^{2n-3}(4n-3-2i)\epsilon _{i}+
(2n-1){{\epsilon _{2n-1}+1} \over 2}$, which agrees with the 
de Rham cohomology of ${\rm SO}(2n)$:$\Lambda (x_{3},x_{7},\cdots,
x_{4n-5},x_{2n-1})$.$^{5}$

\section{${\rm G}_{2}$}
   The group ${\rm G}_{2}$ is a 14-dimensional submanifold of ${\rm O}(7)$.  An element
$A$ of ${\rm G}_{2}$ is given by three 7-components vectors $a_{i}(i=1,2,4)$;
\begin{equation}M(7,3;{\bf R} ) \supset {\rm G}_{2} \ni A =(a_{1},a_{2},a_{4}),
\end{equation}
with conditions $|a_{i}|=1(i=1,2,4)$ and $(a_{1},a_{2})=(a_{1},a_{4})
=(a_{2},a_{4})=(a_{1}a_{2},a_{4})=0$.  The inner product and the 
absolute value are defined by $(a_{i},a_{j})= \displaystyle \sum _{k=1}^{7} a_{ki}a_{kj}$,
$|a_{i}|=\sqrt{(a_{i},a_{i})}$ and the products $a_{i}a_{j}$ are defined
using octonians(see Appendix ).

Fix real numbers $c_{1},c_{2}$ and $c_{4}$ with $c_{1}>c_{2}>c_{4}>0$. Then
\begin{equation}
h=c_{1}a_{11}+c_{2}a_{22}+c_{4}a_{44}
\end{equation}
is a Morse function on ${\rm G}_{2}$.$^{4}$ This Morse function has eight critical 
points with the Morse indices $l=0,3,5,6,8,9,11$ and $14$. We are interested in
the following four classical vacuums $|5 \rangle ,|6 \rangle ,|8 \rangle $ 
and $|9 \rangle $ which are localized around the critical points $(a_{11},a_{22},a_{44})=(-1,1,-1),(1,-1,-1),(-1,1,1)$ and $(1,-1,1)$,
respectively. Around each critical point $(\epsilon _{1},\epsilon _{2},
\epsilon _{4})$,  $h$ has the expansion $^{4}$
\bea
h=\sum _{i} \epsilon _{i}c_{i} +{1 \over 2}
\sum _{i<j}(\lambda _{ij}\xi _{ij}^{2}+
\mu _{ij}\eta _{ij}^{2})-\epsilon _{1}c_{1}(a_{31}^{2}+a_{51}^{2}
+a_{71}^{2})-\epsilon _{2}c_{2}(a_{32}^{2}+a_{62}^{2}+a_{72}^{2}) \nonumber \\ 
-\epsilon _{4}c_{4}(a_{54}^{2}+a_{64}^{2}+a_{74}^{2})
-{1 \over 9}(a_{34} \quad a_{52} \quad a_{61}) B 
(a_{34} \quad a_{52} \quad a_{61})^{t},
\eea
where $i,j=1,2,4$ ,$\lambda _{ij},\mu _{ij}, \xi _{ij}$  and   $\eta _{ij}$
are defined by (27) and (28), and $B$ is the following $3 \times 3$ matrix            
\[ B= \left( 
      \begin{array}{ccc}
  4\epsilon _{4}c_{4}+\epsilon _{2}c_{2}+\epsilon _{1}c_{1}
    &2\epsilon _{2}c_{4}+2\epsilon _{4}c_{2}-\epsilon _{1}\epsilon _{2}
\epsilon _{4}c_{1} &2\epsilon _{1}c_{4}-\epsilon _{1}\epsilon _{2}
\epsilon _{4}c_{2} +2\epsilon _{4}c_{1}       \\
 2\epsilon _{2}c_{4}+2\epsilon _{4}c_{2}-\epsilon _{1}\epsilon _{2}
\epsilon _{4}c_{1}
 & \epsilon _{4}c_{4}+4\epsilon _{2}c_{2}+\epsilon _{1}c_{1}    
& \epsilon _{1}\epsilon _{2}\epsilon _{4}c_{4}- 2\epsilon _{1}c_{2}
-2\epsilon _{2}c_{1}          \\
  2\epsilon _{1}c_{4}-\epsilon _{1}\epsilon _{2}\epsilon _{4}c_{2}
+2\epsilon _{4}c_{1}  & \epsilon _{1}\epsilon _{2}\epsilon _{4}c_{4}
-2\epsilon _{1}c_{2}-2\epsilon _{2}c_{1} 
   &\epsilon _{4}c_{4}+\epsilon _{2}c_{2}+4\epsilon _{1}c_{1} 
      \end{array}
      \right). \]  
 From (39) one finds  classical vacuums as
\bea
|5 \rangle &=&|a_{32},a_{62},a_{72},a_{24}+a_{42},(a_{34},a_{52},a_{61}) \rangle , 
\nonumber \\
|6 \rangle &=&|a_{31},a_{51},a_{71},a_{12}+a_{21},a_{14}+a_{41},
(a_{34},a_{52},a_{61}) \rangle , \nonumber \\
|8 \rangle &=&|a_{32},a_{62},a_{72},a_{24}-a_{42},(a_{34},a_{52},a_{61}),
a_{54},a_{64},a_{74}, \rangle , \nonumber \\
|9 \rangle &=&|a_{31},a_{51},a_{71},a_{12}-a_{21},a_{14}-a_{41},
(a_{34},a_{52},a_{61}),a_{54},a_{64},a_{74} \rangle , 
\eea
where the modes in the ket vectors denote exited fermionic modes and
$(a_{34},a_{52},a_{61})$ represents a certain linear combination of
$ a_{34},a_{52}$ and  $a_{61}$.

Let us discuss instanton solutions of ${\rm G}_{2}$. Embedding ${\rm G}_{2}$ into
${\rm O}(7)$(see Appendix), we see that the critical points that corresponds to
$|5 \rangle ,|6 \rangle ,|8 \rangle $ and $|9 \rangle $ are elements of
${\rm SO}(7)$. The Morse function (38) will be obtained from (24)
by imposing some restrictions on the coordinates for ${\rm SO}(7)$.
Accordingly, instanton solutions of ${\rm G}_{2}$ should be included in that
of ${\rm SO}(7)$. Consider the pair $|5 \rangle $ and $|6 \rangle $.
There are four instanton solutions of ${\rm SO}(7)$ that connect
$ (a_{11},a_{22},a_{44})=(-1,1,-1)$ and $(1,-1,-1)$
(these are diag$(-1,1,-1,-1,1,-1,1)$ and diag$(1,-1,-1,-1,-1,1,1)$ in
${\rm SO}(7)$, respectively): 
\be
\left(
\begin{array}{ccccccc}
\cos \alpha &\pm \sin \alpha & & & & &   \nonumber \\
 \pm \sin \alpha  & -\cos \alpha & & & &  & \nonumber \\
            &   & -1 & & & &  \nonumber  \\
            &   &  & -1 &  &  & \nonumber \\
    &   &  &  & - \cos \beta &\pm \sin \beta &  \nonumber \\
            &   &  &   &     \pm \sin \beta   & \cos \beta   & \nonumber \\
            &   &  &   &    &    & 1    \\
\end{array}
\right),\\
\ee
Setting $\alpha = \beta $ , Eq.(41) represents two paths in ${\rm G}_{2}$
and gives two instanton solutions in ${\rm G}_{2}$. Thus, we have found that
there are exactly two instanton solutions in ${\rm G}_{2}$ that connect 
$|5 \rangle $ and $|6 \rangle $.  Under the 
transformation $a_{12} \rightarrow -a_{12}$,$a_{21} \rightarrow -a_{21}$,
which induces the transformation $a_{56} \rightarrow -a_{56}$,
$a_{65} \rightarrow -a_{65}$ in terms of ${\rm O}(7)$, the two instanton
solutions are interchanged. We use this transformation to elucidate true
vacuums. In the same way, one finds that there are two similar instanton 
solutions between $|8 \rangle $ and $|9 \rangle $.

We will determine whether there are instanton effects or not between the
classical vacuums (40).  A reflection transformation that is useful to this
purpose is such transformation under which the following each set of
modes $(a_{12},
a_{21}),(a_{34},a_{52},a_{61}),$ \\
$(a_{14},a_{41})$ and $(a_{24},a_{42})$ has
the same parities.  For ${\rm SO}(n)$, we have seen that a reflection transformation
that  transform a group element to a group element makes the theory
invariant.$^{1}$  As ${\rm G}_{2}$ is a subgroup of ${\rm O}(7)$, some reflection 
symmetries for ${\rm SO}(n)$ will survive as symmetries for 
${\rm G}_{2}$ with suitable
restrictions.  For ${\rm G}_{2}$, one finds eight reflection transformations
that transform a group element to a group element. These eight transformations
, which are certain combined transformations of $[i]$ for  ${\rm SO}(n)$,
will be symmetries of the theory. Among these eight transformations,
the four transformations in Table I interchange the two instanton solutions
(41) with $\alpha = \beta $.\\
\begin{center}  
           Table I.  \\
\end{center}
The classical vacuums $|5 \rangle $ and $|6 \rangle $ have the same parities
under each of the four transformations. Thus, the matrix element
$\langle 6|d_{h}|5 \rangle $ is non-vanishing, and $|5 \rangle $ and 
$|6 \rangle $ are not true vacuums. In the same way, one finds that
$|8 \rangle $ and $|9 \rangle $ are not true vacuums. We have found that
true vacuums are $|0 \rangle , |3 \rangle ,|11 \rangle$
and $|14 \rangle $ in accordance with de Rham cohomology of ${\rm G}_{2}$; 
$b_{p}=1$ for $p=0,3,11,14$ and $b_{p}=0$ for the others.
\section{${\rm U}(2)$}
In this section we discuss ${\rm U}(2)$, which is topologicaly identical with
${\rm SO}(2) \times {\rm SU}(2)$. We parametrize an element $A$ of ${\rm U}(2)$ as
\bea
{\it A} =\left(  \begin{array}{cc}
x_{11}+iy_{11} & x_{12}+iy_{12}\cr
x_{21}+iy_{21} & x_{22}+iy_{22}
\end{array} \right) =
 \left(  \begin{array}{cc}
\cos \theta e^{i(\alpha +\phi + \psi)} & -\sin \theta e^{i(\alpha +\phi - \psi)} \cr
\sin \theta e^{i(\alpha -\phi + \psi)} & \cos \theta e^{i(\alpha -\phi - \psi)} 
\end{array} \right), \\
 0 \le \theta \le \pi, 0 \le \alpha ,\phi ,\psi \le 2\pi . \nonumber
\eea
Fix real numbers $c_{1},c_{2}$  with $c_{2}>2c_{1}>0$. Then
\begin{equation}
h=c_{1}x_{11}+c_{2}x_{22}= (c_{1} \cos (\alpha +\phi + \psi) 
+ c_{2} \cos (\alpha -\phi - \psi) )\cos \theta
\end{equation}
is a Morse function on ${\rm U}(2)$.$^{4}$ This Morse function has four critical 
points 
\bea
 P^{(0)} = \left( \begin{array}{cc}
-1 & 0 \cr
0 & -1 
\end{array} \right) , 
P^{(1)} = \left( \begin{array}{cc}
1 & 0 \cr
0 & -1 
\end{array} \right) , 
P^{(3)} = \left( \begin{array}{cc}
-1 & 0 \cr
0 & 1 
\end{array} \right) , 
P^{(4)} = \left( \begin{array}{cc}
1 & 0  \cr
0 & 1 
\end{array} \right) .
\eea 
We denote the corresponding classical vacuums as $|0 \rangle ,|1 \rangle ,
|3 \rangle $ and $ |4 \rangle $ , respectively. 

 Around each critical point $(\epsilon _{1},\epsilon _{2})$, $h$ has the expansion 
\be
8h= (\epsilon _{1}c_{1} +\epsilon _{2}c_{2})(8-
(\epsilon_{1}x_{12}-\epsilon_{2}x_{21})^{2}-(\epsilon_{1}y_{12}+\epsilon_{2}y_{21})^{2})-4\epsilon_{1}c_{1}y_{11}^{2}
-4\epsilon_{2}c_{2}y_{22}^{2}.
\ee
From this expansion one  finds that 
\bea
|0 \rangle &=&|0 \rangle ,\qquad |1 \rangle =|\xi \rangle , \\ \nonumber
|3 \rangle &=&|\eta , x_{12}+ x_{21},y_{12}- y_{21} \rangle , \quad 
|4 \rangle = |\xi ,\eta , x_{12}- x_{21},y_{12}+ y_{21} \rangle , 
\eea
where  $\xi =\alpha +\phi + \psi $ and $\eta =\alpha -\phi - \psi $.
We introduce the following metric
\be
g_{ij}={1 \over 2} {\rm tr}{\partial A^{\dagger}  \over \partial {\theta}^{i}}
{\partial A \over \partial {\theta}^{j}},
\ee
where the indices $i$ denote $\theta , \alpha , \psi $ and $ \phi $.
The non-zero components of $g_{ij}$ and Christoffel symbols are
\bea
g_{ii}=1,g_{\phi \psi}=g_{\psi \phi }=\cos 2\theta,  \\ \nonumber 
\Gamma _{\phi \psi}^{\theta}= \Gamma _{\psi \phi }^{\theta}= \sin 2\theta ,
\Gamma _{\theta \psi}^{\phi}= \Gamma _{\theta \phi }^{\psi}
=-{1 \over \sin 2\theta} .
\eea
The covariant derivatives are
\bea
 \nabla _{\alpha}&=&\partial _{\alpha}, \qquad \nabla _{\theta}=\partial _{\theta}
+{1 \over \sin 2\theta}(\hat{\psi}^{* \phi}\hat{\psi}_{ \psi}
+\hat{\psi}^{* \psi}\hat{\psi}_{ \phi}),\\ \nonumber
\nabla _{\phi}&=&\partial_{\phi}-\sin 2\theta\hat{\psi}^{* \psi}
\hat{\psi}_{ \theta}
+{1 \over \sin 2\theta}\hat{\psi}^{* \theta}\hat{\psi}_{ \psi}, \\ \nonumber
\nabla _{\psi}&=&\partial_{\psi}-\sin 2\theta\hat{\psi}^{* \phi}
\hat{\psi}_{ \theta}
+{1 \over \sin 2\theta}\hat{\psi}^{* \theta}\hat{\psi}_{ \phi}.
\eea
Consider the following simultaneous transformation
\be
\theta ^{i} , \hat{\psi}^{* \theta ^{i}},\hat{\psi}_{ \theta ^{i}}
\rightarrow 
-\theta ^{i} , -\hat{\psi}^{* \theta ^{i}},-\hat{\psi}_{ \theta ^{i}},
\theta ^{i}=\alpha ,\psi ,\phi ,
\ee
which makes the Morse function $h$ invariant.
Under this transformation the covariant derivatives 
$ \nabla _{\alpha},\nabla _{\psi}$ and $\nabla _{\phi}$ reverse the sings
and $\nabla _{\theta}$ is invariant. Accordingly, the supercharges
$d_{h},d_{h}^{\dagger}$ and the Hamiltonian is invariant under (50).

The gradient flow equations are
\bea
{d\theta  \over dt} &=&-(c_{1}\cos \xi + c_{2}\cos \eta  )
\sin \theta , \nonumber \\
{d\alpha \over dt} &=&-(c_{1}+c_{2})\sin \theta \sin \alpha 
\cos(\phi+\psi)+(c_{2}-c_{1})\cos \theta \cos \alpha 
\sin(\phi+\psi), \nonumber \\
{d\phi \over dt}&=&{d\psi \over dt} \nonumber \\
&=&((c_{2}-c_{1})\sin \alpha \cos(\phi+\psi)-(c_{1}+c_{2})\cos \alpha \sin(\phi+\psi)){\cos \theta \over {1+\cos 2\theta }}.
\eea
The degrees of the freedom has reduced to three.
Let us consider the equations in the $(\theta , \eta )-\hspace{0pt}$ plane.
There are four fixed points $(\theta , \eta )=(0,0),(0,\pi),(\pi,0)$ and
$(\pi,\pi)$. Around these fixed points the gradient flow equations (51) become
simple:
\bea
\theta \simeq 0 , \qquad {d\eta \over dt} &=&-2c_{2}\sin \eta , \nonumber \\
\theta \simeq \pi , \qquad {d\eta \over dt} &=&2c_{2}\sin \eta , \nonumber \\
\eta \simeq 0 , \qquad {d\theta \over dt} &=&-(c_{2}+c_{1}\cos \xi )
\sin \theta , \nonumber \\
\eta \simeq \pi , \qquad {d\theta \over dt} &=&(c_{2}-c_{1}\cos \xi )
\sin \theta . 
\eea
One sees that $(\theta , \eta )=(0,0),(\pi,\pi)$ are stable fixed
points and that $(\theta , \eta )=(0,\pi),(\pi,0)$ are unstable
fixed points irrespective to the value of $\xi$. 

The initial and the final conditions of a solution
that connects $P^{(0)}$ and $P^{(1)}$ are both 
$\cos \theta \cos \eta = -1$. So, for solutions of (51) that connect  
$P^{(0)}$ and $P^{(1)}$, one concludes that $(\theta , \eta ) \equiv 
(0,\pi)$ or $(\pi,0)$. Under these conditions the gradient flow
equation of $\xi$ becomes
\be
{d\xi \over dt} = \pm 2c_{1}\sin \xi.
\ee
There are exactly two instanton solutions that connect $P^{(0)}$ and $P^{(1)}$
\bea
{\it A} =\left(  \begin{array}{cc}
e^{\pm i\xi} &  0 \cr
0 & -1
\end{array} \right) , \quad  0 \le \xi \le \pi . 
\eea

For solutions that connects  $P^{(3)}$ and $P^{(4)}$, one sees
that $(\theta , \eta ) \equiv (0,0)$ or $(\pi,\pi)$ and that 
there are exactly two instanton solutions 
\bea
{\it A} =\left(  \begin{array}{cc}
e^{\pm i\xi} &  0 \cr
0 & 1
\end{array} \right) , \quad  0 \le \xi \le \pi . 
\eea

By the transformation (50), $\xi $ reverses the sign and the two instanton
solutions (54) ((55)) are interchanged each other. Under this transformation
$|0 \rangle $ and $|3 \rangle $ have even parities and 
$|1 \rangle $ and $|4 \rangle $ have odd parities. Therefor, there
are not instanton effects in the processes  $P^{(0)} \rightarrow P^{(1)}$
and $P^{(3)} \rightarrow P^{(4)}$. Our result for ${\rm U}(2)$ is that 
$|0 \rangle ,|1 \rangle ,|3 \rangle $  and $|4 \rangle $ are all true
vacuums. This result agrees with the de Rham cohomology of ${\rm U}(2)$;
$b_{p}=1$ for $p=0,1,3,4$ and $b_{p}=0$ for the others. Our result is
one of the examples of the Kunneth formula for a product manifold 
$M \times N$ :$b_{k}(M \times N)={\displaystyle \sum} _{p+q=k}
b_{p}(M)b_{q}(N)$.

\section{SUMMARY}
We have studied supersymmetric quantum mechanics
on ${\rm RP}_{n},{\rm SO}(n), {\rm G}_{2}$ and ${\rm U}(2)$ as 
concrete examples of Witten's Morse theory.
We have identified all the true vacuums for each theory. 
The number of the true vacuums agrees with the de Rham
cohomology for each manifold.  We have seen that the simple instanton
picture that the author proposed for $SO(n)^{4}$ holds for the above
manifolds.  For each pair of adjacent classical vacuums,
there are exactly two instanton solutions which are 
interchanged by a reflection transformation
of the theory. Owing to the reflection symmetry, it is easy to 
decide whether the two instanton effects cancel each other or not. 
\\
{\bf ACKNOWLEDGMENTS}  \\
The author thanks Dr.Y.Yasui for introducing him to this subject.
The author thanks Dr.H.Usui for useful discussions.

\section*{APPENDIX:MULTIPLICATION RULE  in ${\rm G}_{2}$ and EMBEDDING}
 We note the multiplication rule in ${\rm G}_{2}$. For 
elements $a_{i}$ of ${\rm G}_{2}$, the following expansions by octonians $e_{j}$
are taken into account
\[ a_{i} = \sum _{j=1}^{7}a_{ji}e_{j}=  \left(
           \begin{array}{@{\,}c@{\,}}
                a_{1i} \\
                \vdots \\
                a_{7i}
           \end{array}
           \right). \]
Note that $e_{0}=1$ does not appear in the expansion. 
\\
\\
Table. II.
\\
\\
The multiplication rules of octonians are defined as usual(see Table II);
for example,
$e_{1}e_{2}=e_{3}$,$e_{5}e_{3}=-e_{6}$,and $e_{i}^{2}=-1$,
$e_{i}e_{j}=-e_{j}e_{i}(i \neq j)$. Explicitly, $a_{3}=a_{1}a_{2}$ is
given by
\[ a_{3} =  \left(
           \begin{array}{@{\,}c@{\,}}
-a_{31}a_{22}+a_{21}a_{32}-a_{51}a_{42}+a_{41}a_{52}-a_{71}a_{62}+a_{61}a_{72}\\
a_{31}a_{12}-a_{11}a_{32}+a_{61}a_{42}-a_{41}a_{62}-a_{71}a_{52}+a_{51}a_{72}\\
-a_{21}a_{12}+a_{11}a_{22}-a_{71}a_{42}-a_{61}a_{52}+a_{51}a_{62}+a_{41}a_{72}                      \\
a_{51}a_{12}-a_{61}a_{22}+a_{71}a_{32}-a_{11}a_{52}+a_{21}a_{62}-a_{31}a_{72}                 \\
-a_{41}a_{12}+a_{71}a_{22}+a_{61}a_{32}+a_{11}a_{42}-a_{31}a_{62}-a_{21}a_{72}                        \\
a_{71}a_{12}+a_{41}a_{22}+a_{51}a_{32}-a_{21}a_{42}-a_{31}a_{52}-a_{11}a_{72}                       \\
-a_{61}a_{12}-a_{51}a_{22}-a_{41}a_{32}+a_{31}a_{42}+a_{21}a_{52}+a_{11}a_{62}                       \\
\end{array}
           \right). \]

An element $A$ of ${\rm G}_{2}$ is embedded into ${\rm O}(7)$ as follows $^{4}$
\\
$
A=(a_{1},a_{2},a_{4}) \rightarrow (a_{1},a_{2},a_{1}a_{2},a_{4},a_{1}a_{4},a_{4}a_{2},a_{1}a_{4}a_{2})
\in {\rm O}(7). 
$
\\
\newpage
{\bf References}\\
$^1$K.Oshima, { J.Math.Phys.}{\bf 38},6281(1997). \\
$^2$E.Witten,{ J.Diff.Geom.}{\bf 17},661(1982). \\
$^3$T.Hirokane,M.Miyajima and Y.Yasui,{ J.Math.Phys.}{\bf 34},2789(1993); \\
\quad K.Oshima, Prog.Theor.Phys.{\bf 96},1237(1996). \\
$^4$I.Yokota,{\it Manifold and Morse Theory(in Japanese)}(Gendai Suugakusha,Kyoto,1978).
$^5$S.Iyanaga and Y.Kawada ,{\it Encyclopedic Dictionary
of Mathematics}(The MIT Press, Cambridge,1977).\\
\newpage
TABLE I. Symmetry transformations that interchange the two instanton
solutions (41) with $\alpha = \beta $. 
The minus signs mean that the corresponding elements  $a_{ki}$
,$i=1, \cdots ,4, k=1,\cdots ,7,$ reverse the signs($a_{3}$ is not independent).\\ 
\\
{\footnotesize
\begin{tabular}{|c|cccc|} \hline
   & 1 &  2 & 3  & 4  \\ \hline      
1  & + & $-$ &$ -$ & +  \\ 
2  & $-$ & + & + & $-$   \\  
3  & $- $& + & + & $-$   \\ 
4  & + & $-$ & $-$ & +   \\ 
5  & + &$ -$ &$ -$ & +   \\ 
6  & $-$ & + & + & $-$   \\ 
7  & $-$ & + & + &$ -$   \\ \hline
\end{tabular} \quad
\begin{tabular}{|c|cccc|} \hline
   & 1 &  2 & 3  & 4  \\ \hline      
1  & + & $-$ & + & +  \\ 
2  & $-$ & + & $-$ &$ -$   \\  
3  & + & $-$ & + & +   \\ 
4  & + & $-$ & + & +   \\ 
5  & $-$ & + & $-$ & $-$   \\ 
6  & + & $-$ & + & +   \\ 
7  & $-$ & + & $-$ & $-$   \\ \hline
\end{tabular} \quad
\begin{tabular}{|c|cccc|} \hline
   & 1 &  2 & 3  & 4  \\ \hline      
1  & + & $-$ & + & $-$  \\ 
2  & $-$ & + & $-$ & +   \\  
3  & + & $-$ & + & $-$   \\ 
4  & $-$ & + & $-$ & +   \\ 
5  & + & $-$ & + & $-$   \\ 
6  & $-$ & + & $-$ & +   \\ 
7  & + & $-$ & + & $-$   \\ \hline
\end{tabular} \quad
\begin{tabular}{|c|cccc|} \hline
   & 1 &  2 & 3  & 4  \\ \hline      
1  & + & $-$ & $-$ & $-$  \\ 
2  & $-$ & + & + & +   \\  
3  & $-$ & + & + & +   \\ 
4  & $-$ & + & + & +   \\ 
5  & $-$ & + & + & +   \\ 
6  & + & $-$ & $-$ &$ -$   \\ 
7  & + & $-$ &$ -$ & $-$   \\ \hline
\end{tabular}
}
\\
\\
\\
TABLE II. Multiplication table of octonians $e _{j}$.\\ 
\\
\begin{tabular}{|c|ccccccc|} \hline
     & $e_{1}$ & $e_{2}$ & $e_{3}$&$ e_{4}$ &$ e_{5}$& $e_{6}$ & $e_{7}$ \\ \hline      
$e_{1}$&$   -1 $ & $e_{3}$ &$-e_{2}$&$ e_{5}$ &$ -e_{4}$& $e_{7}$&$ -e_{6}$  \\ 
$e_{2}$&$ -e_{3}$& $-1$    & $e_{1}$&$ -e_{6}$ & $e_{7}$& $e_{4}$& $-e_{5}$   \\  
$e_{3}$& $ e_{2}$& $-e_{1}$&  $  -1$& $ e_{7}$ &$ e_{6}$& $-e_{5}$&$ -e_{4}$  \\ 
$e_{4}$& $-e_{5}$& $e_{6}$& $ -e_{7}$& $-1 $  & $e_{1}$&$ -e_{2}$&$ e_{3}$ \\ 
$e_{5}$& $e_{4}$& $-e_{7}$& $ -e_{6}$&$ -e_{1}$ & $- 1$& $ e_{3}$& $e_{2}$  \\ 
$e_{6}$& $-e_{7}$& $-e_{4}$& $ e_{5}$& $e_{2}$ & $-e_{3}$& $  -1$& $e_{1}$  \\ 
$e_{7}$& $ e_{6}$& $e_{5}$& $ e_{4}$& $-e_{3}$ & $-e_{2}$&$ -e_{1}$& $-1$   \\ \hline
\end{tabular} 
\end{document}